\documentclass[journal]{IEEEtran}
\usepackage{amsfonts}
\usepackage{cite}
\usepackage{graphicx}
\usepackage{amsmath}
\usepackage{amssymb}
\usepackage{algorithm}
\usepackage{algorithmic}

\begin{document}

\title{On the BCJR Algorithm for Asynchronous Physical-layer Network Coding}

\author{Xiaofu~Wu, ~Chunming~Zhao,
        and Xiaohu~You 
\thanks{This work was supported in part by the National Science
        Foundation of China under Grants 60972060, 61032004. The work of X. Wu was also supported by the National Key S\&T Project under Grant 2010ZX03003-003-01 and by the Open Research Fund of National Mobile Communications Research Laboratory, Southeast University (No. 2010D03).
        }
\thanks{Xiaofu Wu is with the Nanjing Institute of Communications
        Engineering, PLA Univ. of Sci.\&Tech., Nanjing 210007, China. He is also with the National Mobile Commun. Research Lab., Southeast Univ., Nanjing
        210096 (Email: xfuwu@ieee.org).}
\thanks{Chunming Zhao and Xiaohu You are with the National Mobile Commun. Research Lab.,
        Southeast University, Nanjing 210096, China (Email: cmzhao@seu.edu.cn, xhyu@seu.edu.cn).}}


\maketitle

\begin{abstract}
In practical asynchronous bi-directional relaying, symbols transmitted by two source nodes cannot arrive at the relay with perfect symbol alignment and the symbol-asynchronous multiple-access channel (MAC) should be seriously considered. Recently, Lu et al. proposed a Tanner-graph representation of symbol-asynchronous MAC with rectangular-pulse shaping and further developed the message-passing algorithm for optimal decoding of the asynchronous physical-layer network coding.  In this paper, we present a general channel model for the asynchronous multiple-access channel with arbitrary pulse-shaping. Then, the Bahl, Cocke, Jelinek, and Raviv (BCJR) algorithm is developed for optimal decoding of asynchronous MAC channel. This formulation can be well employed to develop various low-complexity algorithms, such as Log-MAP algorithm, Max-Log-MAP algorithm, which are favorable in practice.
\end{abstract}


\IEEEpeerreviewmaketitle

\section{Introduction}
\PARstart{N}{etwork} coding has shown its power for disseminating information over networks \cite{KoetterAlg,Chou}.
For wireless cooperative networks, there are increased interests in employing the idea of network
coding for improving the throughput of the network.
Indeed, the gain is very impressive for the special bi-directional relaying
scenarios with two-way or multi-way traffic as addressed in \cite{Popovski}.

For bi-directional relaying, it was soon recognized that the superimposed signal received at the relay can be viewed as the physically-combined network coding form of the two source messages, which is further impaired by the channel noise. Hence,  the so-called physical-layer network coding (PNC) \cite{Zhang_PLNC} can be well employed to improve the throughput of the bi-directional relaying.

For bi-directional relaying with PNC,  it is assumed that communication takes place in two phases - a multiple
access (MAC) phase and a broadcast phase. In the first phase, the two source nodes send signals simultaneously to the relay;
in the second phase, the relay processes the superimposed of the simultaneous packets and
maps them to a network-coded packet for broadcast back to the source nodes. Compared with the traditional relay system, PNC doubles the throughput of
the two-way relay channel by reducing the time slots for the exchange of one packet from four to two.

A key issue in practical PNC is how to deal with the asynchrony between the signals transmitted
by the two source nodes. That is, symbols transmitted by the two source nodes
could arrive at the receiver with symbol misalignment.

In \cite{LuWeb}, Lu et al. proposed a Tanner-graph representation of symbol-asynchronous multiple-access channel (MAC) with rectangular-pulse shaping  and further developed the message-passing algorithm for optimal decoding of asynchronous physical-layer network coding.

In this paper, we provide further insights into the optimal decoding for asynchronous physical-layer network coding. In particular, the general asynchronous MAC channel with arbitrary pulse-shaping is developed and its connection to the rectangular-pulse shaping \cite{LuWeb} is discussed. Then, the BCJR formulation of the asynchronous MAC channel is proposed, which can shed lights for various practical algorithms suitable for implementation.

\section{General Channel Model for Asynchronous Physical-layer Network Coding}
\subsection{Asynchronous Multiple Access Channel Model}

During the MAC phase, the source nodes A and B transmit the modulated signals $x_a(t)$ and $x_b(t)$ to the relay.
For a general continues-time multiple-access channel, the received signal at the relay can be
expressed as
\begin{eqnarray}
 \label{eq:c1}
    y(t) &=& h_a x_a(t)+ h_b x_b(t)+w(t) \nonumber \\
           &=& \sum_{k=1}^{\infty} h_a c_a(k)g_a(t-k T - \tau_a) \nonumber \\
              & &+ \sum_{k=1}^{\infty} h_b c_b(k) g_b(t-k T - \tau_b) + w(t),
\end{eqnarray}
where the delays $\tau_a \in [0,T), \tau_b \in [0,T)$ account for the symbol asynchronism between source nodes A and B and known to the receiver,  $w(t)$ is the complex white Gaussian noise with power spectral density equal to $\frac{\sigma^2}{2}$, the channel coefficients $h_a, h_b$ are complex channel gains keeping fixed during transmission, and $g_a(t), g_b(t)$ are normalized pulse-shaping functions ($\frac{1}{T} \int_0^T |g_a(t)|^2 dt = 1$) for source nodes A and B, respectively. Without loss of generality, we assume that $0\leq \tau_a \leq \tau_b<T$.

By passing the observations through two matched filters for signals ($x_a(t)$) and ($x_b(t)$), respectively, one can get the following discrete-time samples
\begin{eqnarray}
 \label{eq:d1}
    y_a(k) = \frac{1}{T} \int_{kT+\tau_a}^{(k+1)T+\tau_a} y(t)g_a^*(t-kT-\tau_a)dt, \nonumber \\
    y_b(k) = \frac{1}{T} \int_{kT+\tau_b}^{(k+1)T+\tau_b} y(t)g_b^*(t-kT-\tau_b)dt.
\end{eqnarray}

It was recognized that the discrete samples $\left\{[y_a(k), y_b(k)]^T\right\}$ are sufficient statistics for maximum a posteriori (MAP) symbol detection as explained in \cite{VerduIT89}. By incorporating (\ref{eq:c1}) into (\ref{eq:d1}), it follows that
\begin{eqnarray}
 \label{eq:d2}
    y_a(k) = h_b \rho_{ba}^* c_b(k-1) + h_a c_a(k) + h_b \rho_{ab} c_b(k) + w_a(k), \nonumber \\
    y_b(k) = h_a \rho_{ab}^* c_a(k) + h_b c_b(k) + h_a \rho_{ba} c_a(k+1) + w_b(k),
\end{eqnarray}
where
\begin{eqnarray}
 \label{eq:d3}
    \rho_{ab} = \frac{1}{T} \int_0^T g_a^*(t) g_b(t+\tau_a-\tau_b) dt,
\end{eqnarray}
and
\begin{eqnarray}
 \label{eq:d4}
    \rho_{ba} = \frac{1}{T} \int_0^T g_a(t) g_b^*(t+T+\tau_a-\tau_b) dt.
\end{eqnarray}

One can also rewrite (\ref{eq:d2}) in the matrix form, as shown in the top of the next page.
\begin{figure*}[!t]
\begin{eqnarray}
\label{eq:A3}
\left[\begin{array}{c}
y_a(k) \\
y_b(k)
\end{array}\right] = \left[\begin{array}{cc}
0 & h_b\rho_{ba}^* \\
0 & 0
\end{array}\right]
\left[\begin{array}{c}
c_a(k-1) \\
c_b(k-1)
\end{array}\right] +  \left[\begin{array}{cc}
h_a & h_b\rho_{ab} \\
h_a\rho_{ab}^* & h_b
\end{array}\right]
\left[\begin{array}{c}
c_a(k) \\
c_b(k)
\end{array}\right] 
 +  \left[\begin{array}{cc}
0 & 0 \\
h_a \rho_{ba} & 0
\end{array}\right]
\left[\begin{array}{c}
c_a(k+1) \\
c_b(k+1)
\end{array}\right] +
\left[\begin{array}{c}
w_a(k) \\
w_b(k)
\end{array}\right]
\end{eqnarray}
\centering
\end{figure*}
Here, the discrete random process $\left\{[w_a(k),w_b(k)]^T\right\}$ is Gaussian with zero mean and covariance matrix:
\begin{eqnarray}
\label{eq:noise1}
E\left[\left[\begin{array}{c}
w_a(k) \\
w_b(k)
\end{array}\right] \cdot
\left[\begin{array}{cc}
w_a^*(l), w_b^*(l)
\end{array}\right]\right] = \sigma^2 \mathbf{\Lambda}(k-l)
\end{eqnarray}
where $\mathbf{\Lambda}(k)=0$ if $|k|>1$ and $\mathbf{\Lambda}(0)$, $\mathbf{\Lambda}(1)$, $\mathbf{\Lambda}(-1)$ are given as follows
\begin{eqnarray}
    \label{eq:cor0}
    \mathbf{\Lambda}(0) = \left[\begin{array}{cc}
    1 & \rho_{ab} \\
    \rho_{ab}^* & 1
\end{array}\right],
\end{eqnarray}
\begin{eqnarray}
    \label{eqn:cor1}
    \mathbf{\Lambda}(1) = \mathbf{\Lambda}(-1)^\dag = \left[\begin{array}{cc}
    0 & \rho_{ba} \\
    0 & 0
\end{array}\right].
\end{eqnarray}

For convenience of the MAP detection, the whitened matched filter (WMF) is often employed for transforming the received signal into a discrete
time sequence with minimum-phase channel response and white noise. This procedure often simplifies analysis and is a first step in the implementation of some estimators, including the maximum-likelihood sequence estimation
detector (MLSE) and the MAP detector. The WMF is determined by factoring the channel
spectrum into a product of a minimum phase filter and its time inverse.

Let $\mathbf{\Omega}(z)$ be the (two-sized) $z$ transform of the sampled autocorrelation sequences $\mathbf{\Lambda}(k)$, i.e.,
\begin{eqnarray}
    \label{eqn:zf4}
    \mathbf{\Omega}(z) = \sum_k \mathbf{\Lambda}(k)z^{-k} = \mathbf{\Lambda}(-1)z +  \mathbf{\Lambda}(0) + \mathbf{\Lambda}(1)z^{-1}.
\end{eqnarray}
By noting the property (\ref{eqn:cor1}), it follows that ${\Omega}(z)$ can be factored as
\begin{eqnarray}
    \label{eqn:factor}
    \mathbf{\Omega}(z) = \mathbf{F}^\dag(z^{-1}) \mathbf{F}(z).
\end{eqnarray}
By invoking the spectral factorization theorem, it is reasonable to find a physically realizable, stable discrete-time filter $\left(\mathbf{F}^\dag(z^{-1})\right)^{-1}$, which can transform a colored random process into a white random process.

In \cite{Duel}, it has been proved that  $\mathbf{F}(z)$ has the form of
\begin{eqnarray}
    \label{eqn:f1}
    \mathbf{F}(z) = \left[\begin{array}{cc}
    f_{aa} & f_{ab} z^{-1} \\
    f_{ba} & f_{bb}
\end{array}\right].
\end{eqnarray}

Consequently, passage of the received vector sequence $\left\{\mathbf{y}(k)=[y_a(k),y_b(k)]^T\right\}$ through the digital filter $\left(\mathbf{F}^\dag(z^{-1})\right)^{-1}$ results into an output vector sequence $\left\{\mathbf{r}(k)\right\}$ that can be expressed at the top of the next page. Now, the discrete random process $\left\{\mathbf{n}(k)=[n_a(k),n_b(k)]^T\right\}$ is  zero-mean white Gaussian process with covariance of $\sigma^2 \mathbf{I}$.
\begin{figure*}[!t]
\begin{eqnarray}
\label{eq:W}
\left[\begin{array}{c}
r_a(k) \\
r_b(k)
\end{array}\right] = \left[\begin{array}{cc}
0 & h_b f_{ab} \\
0 & 0
\end{array}\right]
\left[\begin{array}{c}
c_a(k-1) \\
c_b(k-1)
\end{array}\right] +  \left[\begin{array}{cc}
h_a f_{aa} & 0 \\
h_a f_{ba} & h_b f_{bb}
\end{array}\right]
\left[\begin{array}{c}
c_a(k) \\
c_b(k)
\end{array}\right] 
 +
\left[\begin{array}{c}
n_a(k) \\
n_b(k)
\end{array}\right]
\end{eqnarray}
\centering
\end{figure*}

Let $\mathbf{c}_{ab}(k)=\left[\begin{array}{c}
c_a(k) \\
c_b(k)
\end{array}\right]$, and $\mathbf{r}(k)=\left[\begin{array}{c}
r_a(k) \\
r_b(k)
\end{array}\right]$. Then, the formulation (\ref{eq:W}) can be elegantly expressed as
\begin{eqnarray}
 \label{eq:d4}
    \mathbf{r}(k) &=& \Psi \left(\mathbf{c}_{ab}(k), \mathbf{c}_{ab}(k-1)\right) + \mathbf{n}(k).
\end{eqnarray}
It is clear that the function $\Psi(\cdot,\cdot)$ is linear. By assuming the ideal knowledge on $\Psi(\cdot,\cdot)$, $\sigma^2$, the asynchronous MAC channel can be modeled as the vector inter-symbol interference (ISI) channel. To estimate the a $posteriori$ probability (APP) $\Pr\left(\mathbf{c}_{ab}(k)|\mathbf{y}_0^{N-1}\right)$, the BCJR algorithm can be naturally employed.

\subsection{Rectangular-pulse shaping}
Let $\delta = \frac{\tau_b-\tau_a}{T}$ denote the relative delay between source nodes A and B. For the rectangular pulse-shaping functions $g_a(t), g_b(t)$, i.e., $g_a(t)=g_b(t)=u(t)-u(t-T)$ with $u(t)$ denoting the unit step function, the authors in \cite{LuWeb} proposed to consider the following discrete-time samples
\setlength{\arraycolsep}{0.0em}
\begin{eqnarray}
 \label{eq:d1}
    y_e(k) &=& \frac{1}{\delta T} \int_{kT+\tau_a}^{kT+\tau_b} y(t)g(t-kT-\tau_a)dt  \nonumber \\
    y_o(k) &=& \frac{1}{(1-\delta)T} \int_{kT+\tau_b}^{(k+1)T+\tau_a} y(t)g(t-kT-\tau_a)dt.
\end{eqnarray}
It is clear that $y_a(k)=\delta y_e(k)+ (1-\delta) y_o(k), y_b(k)=(1-\delta)y_o(k)+\delta y_e(k+1)$. Hence, the samples $\left\{[y_e(k), y_o(k)]^T\right\}$ are also the sufficient statistics for MAP detection.
By combining (\ref{eq:c1}) and (\ref{eq:d1}), it follows that
\begin{eqnarray}
 \label{eq:d2}
    y_e(k) &=& h_a c_a(k) + h_b c_b(k-1) + w_e(k) \nonumber \\
    y_o(k) &=& h_a c_a(k) + h_b c_b(k) + w_o(k),
\end{eqnarray}
\setlength{\arraycolsep}{5pt}
where $w_e(k)$ and $w_o(k)$ are independent zero-mean complex Gaussian variables with variance of $\frac{1}{\delta} \sigma^2$ and $\frac{1}{1-\delta} \sigma^2$. Hence, one can write (\ref{eq:d2}) as the following matrix form

\begin{eqnarray}
\label{eq:d3}
\left[\begin{array}{c}
y_e(k) \\
y_o(k)
\end{array}\right] &=& \left[\begin{array}{cc}
0 & h_b \\
0 & 0
\end{array}\right]
\left[\begin{array}{c}
c_a(k-1) \\
c_b(k-1)
\end{array}\right] \nonumber \\
&+&  \left[\begin{array}{cc}
h_a & 0 \\
h_a & h_b
\end{array}\right]
\left[\begin{array}{c}
c_a(k) \\
c_b(k)
\end{array}\right]
+
\left[\begin{array}{c}
w_e(k) \\
w_o(k)
\end{array}\right].
\end{eqnarray}
Hence, the equivalent ISI channel model (\ref{eq:d4}) is still valid.

\section{BCJR Algorithm}
In this section, we formulate the BCJR algorithm \cite{BCJR}, which is known to be optimal in implementing the
MAP symbol detection for channels with finite memory.

Let us define, at time epoch $k$, the state $s_k$ as
\begin{equation}
\label{eq:sd}
   s_k=\left(\mathbf{c}_{ab}(k-1)\right) = \left(c_a(k-1), c_b(k-1)\right)
\end{equation}
and the branch metric function as
\setlength{\arraycolsep}{0pt}
\begin{eqnarray}
\label{eq:sd}
   \gamma_k && (s_k,\mathbf{c}_{ab}(k))  \nonumber   \\
    && \propto \exp\left(-\frac{\left|\mathbf{r}(k)- \Psi \left(\mathbf{c}_{ab}(k), \mathbf{c}_{ab}(k-1)\right)\right|^2}{2\sigma^2}\right).
\end{eqnarray}
\setlength{\arraycolsep}{5pt}
The BCJR algorithm is characterized by the following forward
and backward recursions:
\begin{eqnarray}
\label{eq:fd}
   \alpha_{k+1}(s_{k+1})=\sum_{\mathbf{c}_{ab}(k)} \sum_{s_k} \mathcal{T}(\mathbf{c}_{ab}(k),s_k,s_{k+1}) \nonumber \\
    \cdot \alpha_k(s_k) \gamma_k(s_k,\mathbf{c}_{ab}(k)),
\end{eqnarray}
where $\mathcal{T}(\mathbf{c}_{ab}(k),s_k,s_{k+1})$ is the trellis indicator function, which is equal to 1 if $\mathbf{c}_{ab}(k), s_k, s_{k+1}$
satisfy the trellis constraint and 0 otherwise;

\begin{eqnarray}
\label{eq:fd}
   \beta_{k}(s_k)=\sum_{\mathbf{c}_{ab}(k)} \sum_{s_{k+1}} \mathcal{T}(\mathbf{c}_{ab}(k),s_k,s_{k+1}) \nonumber \\
   \cdot \beta_{k+1}(s_{k+1}) \gamma_k(s_k,\mathbf{c}_{ab}(k)).
\end{eqnarray}
Then, the joint APPs $\Pr\left(\mathbf{c}_{ab}(k)|\mathbf{y}_0^{N-1}\right)$ can be calculated as
\setlength{\arraycolsep}{0.0em}
\begin{eqnarray}
\label{eq:llr}
    \Pr &&\left(\mathbf{c}_{ab}(k)|\mathbf{y}_0^{N-1}\right)  \nonumber \\
    && =\sum_{s_{k+1}}\mathcal{T}(\mathbf{c}_{ab}(k),s_{k+1}) \alpha_{k+1}(s_{k+1}) \beta_{k+1}(s_{k+1}),
\end{eqnarray}
where the indicator function $\mathcal{T}(\mathbf{c}_{ab}(k),s_{k+1})$ is equal to 1 if $s_{k+1}$
is compatible with $\mathbf{c}_{ab}(k)$ and 0 otherwise.

It should be pointed out that the value of $\Psi \left(\mathbf{c}_{ab}(k), \mathbf{c}_{ab}(k-1)\right)$ is independent of $c_a(k-1)$, hence the state $s_k$ can be further simplified as $s_k= \left( c_b(k-1) \right)$.

Just like in \cite{Hagenauer}, the proposed BCJR algorithm can be implemented efficiently in Log-domain, i.e., Log-MAP algorithm. The further simplification to the Max-Log-MAP algorithm is also straightforward, with some potential performance loss.

With the joint APPs $\Pr\left(\mathbf{c}_{ab}(k)|\mathbf{y}_0^{N-1}\right)$, one can calculate the APPs of the XOR codeword  $\Pr\left(c_a(k)\oplus c_b(k)|\mathbf{y}_0^{N-1}\right)$ for physical network coding. If both source A and B assumes the same linear channel coding, the relay node can make use of $\Pr\left(c_a(k)\oplus c_b(k)|\mathbf{y}_0^{N-1}\right)$ to perform channel decoding to obtain the pairwise XOR of the source symbols. However, this disjoint channel-decoding and network-coding scheme performs worse than the joint channel-decoding and network-coding scheme\cite{Wubben, LuWeb}.

\section{Conclusion and Future Work}
We have presented a general channel model for the asynchronous multiple-access channel with arbitrary pulse-shaping, typically encountered in bi-directional relaying. By evoking the WMF technique,  one can arrive at an equivalent vector ISI channel, which can be employed to develop the well-known BCJR algorithm for getting the optimal APPs. This formulation can be well employed to develop various low-complexity algorithms, such as Log-MAP algorithm, Max-Log-MAP algorithm, which are favorable in practice.

Channel coding can be well employed to improve the system performance. It has been reported in \cite{Wubben, LuWeb} that the joint channel-decoding and network-coding scheme can perform better than the disjoint channel-decoding and network-coding scheme. For joint network and LDPC coding over the asynchronous bi-directional relaying, it is interesting to find more efficient log-domain decoding algorithms suitable for practical implementation.

\end{document}